\documentclass[preprint,showpacs,preprintnumbers,amsmath,amssymb,superscriptaddress]{revtex4-1}
\usepackage{graphicx}
\usepackage{dcolumn}
\usepackage{bm}
\usepackage{color}
\begin{document}
\title{Photoexcitation of a polarization-inverted domain from the charge-ordered 
ferroelectric ground state of (TMTTF)$_{2}$PF$_{6}$}
\author{T. Yamaguchi}
\affiliation{Institute of Materials Structure Science,
High Energy Accelerator Research Organization (KEK), 1-1 Oho, Tsukuba 305-0801, Japan}
\author{K. Asada}
\affiliation{Department of Advanced Materials Science, The University of Tokyo, 5-1-5 Kashiwa-no-ha, Chiba 277-8561, Japan}
\author{H. Yamakawa}
\affiliation{Department of Advanced Materials Science, The University of Tokyo, 5-1-5 Kashiwa-no-ha, Chiba 277-8561, Japan}
\author{T. Miyamoto}
\affiliation{Department of Advanced Materials Science, The University of Tokyo, 5-1-5 Kashiwa-no-ha, Chiba 277-8561, Japan}
\author{K. Iwano}
\affiliation{Graduate University for Advanced Studies, Institute of Materials Structure Science,
High Energy Accelerator Research Organization (KEK), 1-1 Oho, Tsukuba 305-0801, Japan}
\author{T. Nakamura}
\affiliation{Institute for Molecular Science, Myodaiji, Okazaki, 444-8585, Japan}
\author{N. Kida}
\affiliation{Department of Advanced Materials Science, The University of Tokyo, 5-1-5 Kashiwa-no-ha, Chiba 277-8561, Japan}
\author{H. Okamoto}
\affiliation{Department of Advanced Materials Science, The University of Tokyo, 5-1-5 Kashiwa-no-ha, Chiba 277-8561, Japan}
\affiliation{AIST-UTokyo Advanced Operando-Measurement Technology Open Innovation Laboratory,
National Institute of Advanced Industrial Science and Technology, Chiba 277-8568, Japan}
\date{\today}
%
%
%
\begin{abstract}
We theoretically revealed that a weak photoexcitation achieves the electric polarization-inversion with approximately 18$\%$ of 
all the charges, which was interpreted as a superimposition of multi-exciton states, from the charge-ordered ferroelectric 
ground state of (TMTTF)$_{2}$PF$_{6}$ at absolute zero temperature. 
Regarding a relative change of electric polarization ($\Delta P/P$), the photoexcitation corresponds to 36$\%$, 
which is much larger than $\Delta P/P$ of other typical organic materials. The value of $\Delta P/P\sim 36\%$ can be enlarged 
by a strong photoexcitation. This fact is useful not only for applications of this material and other analogous materials 
in optical devices but also for researches toward controlling electric polarizations by light, which is one of the recent 
attracting issues on photoinduced phase transition phenomena. The photoexcitation of $\Delta P/P\sim 36\%$ 
corresponds to the single peak of the optical conductivity in the low-energy region, which was also observed at 10 K. 
Theoretical calculations are based on a quarter-filled one-dimensional effective model with appropriate 
parameters and 50 unit cells. 
\end{abstract}
\maketitle
%
%
\section{Introduction}
%
%
\par
\begin{figure}[t]
\begin{center}
\includegraphics[width=10cm,keepaspectratio]{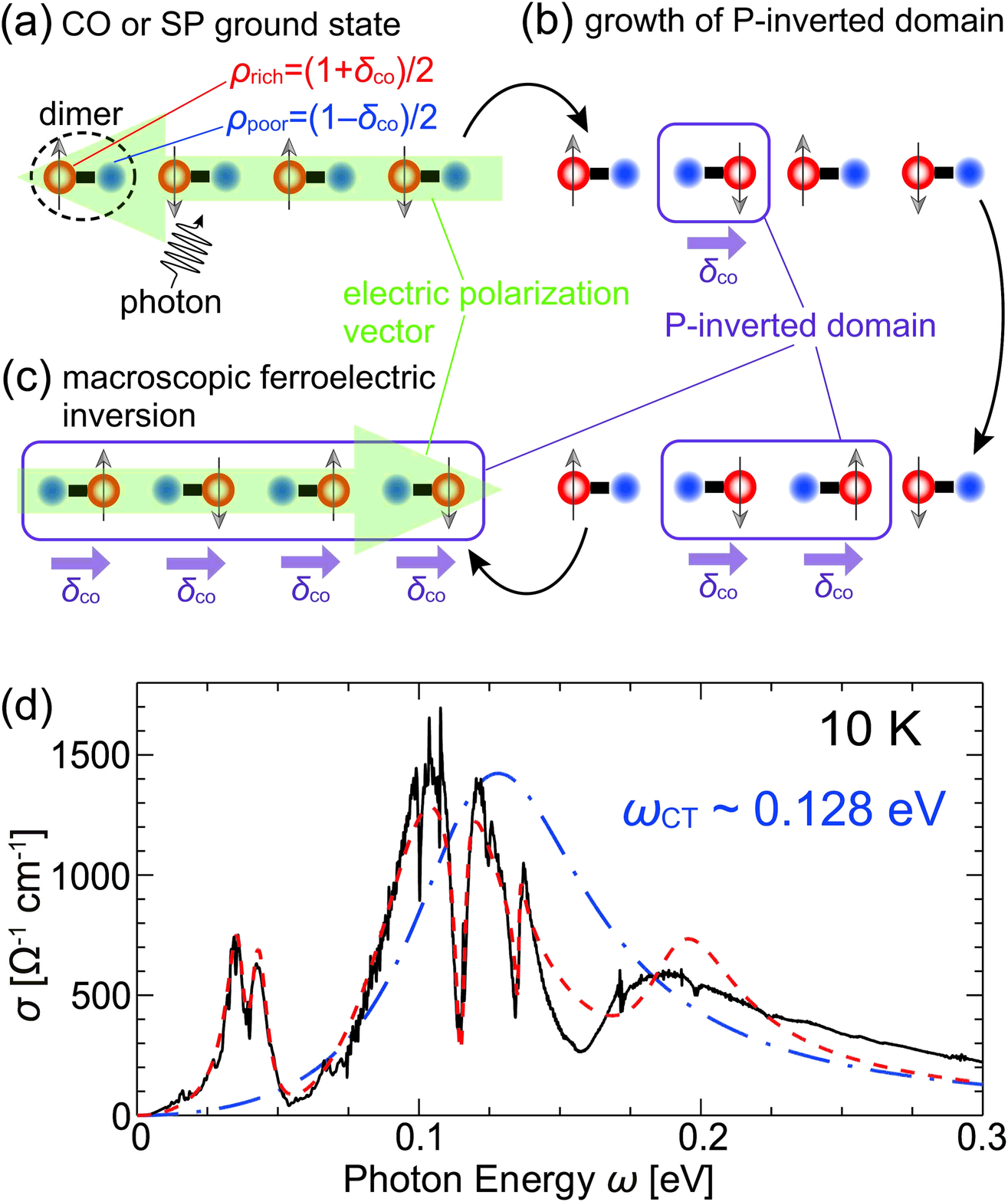}
\caption{(a)--(c) Schematics of the CO or SP ground state and photoexcited states of (TMTTF)$_{2}$PF$_{6}$. 
The circle and up (down) arrows on the circles represent a molecular orbital of a TMTTF molecule and up (down) spins, respectively. 
(d) Optical conductivity spectrum of (TMTTF)$_{2}$PF$_{6}$ with the electric field polarized parallel to the $a$-axis at 10 K 
(solid line) and the fitting curve (dashed line). The chained line shows the calculated spectrum with resonant energy 0.128 eV. 
}
\label{fig1}
\end{center}
\end{figure}
\par
\par
Studies on controlling the purely electronic phase transitions occurring immediately after a photoexcitation from the ground state of 
matter have been attracting attention because such photoinduced phase transitions (PIPTs) regulate the macroscopic properties of 
matter on the ultrafast time scale \cite{PIPT1,PIPT2}. Once such electronic PIPTs are applied to organic 
ferroelectric materials, the electric polarization can be tuned in the regime of femtoseconds. 
Because of notable properties such as mechanical flexibility, disposability, and inexpensiveness, organic materials are increasingly 
being applied to electronic and optical devices. In this regard, flexible tuning of a light-induced electric polarization 
in the order of femtoseconds is one of the most attracting challenge in the field of PIPTs, recently. 
\par
So far, as one of the light sources to easily create and control such devices, visible-light (light) is 
actually most convenient. In this regard, PIPTs induced by light has actively been studied. For instance, 
a light-induced ultrafast insulator--metal transition has been observed in a quasi-two-dimensional molecular solid, 
$\alpha$-(BEDT-TTF)$_2$I$_3$ (BEDT-TTF = bis(ethylenedithiolo)tetrathiafulvalene) \cite{iwai}. Because the material 
undergoes ferroelectric polarization in the charge-ordered (CO) ground state \cite{yamamoto,yamakawa}, this transition is regarded 
as a photoinduced disappearance of the polarization. A photoexcitation of a non-polarized state from a ferroelectric polarized 
ground state has been reported for a quasi-one-dimensional (1D) molecular solid, 
TTF-CA (tetrathiafulvalene-chloranil) \cite{PIPT5}. Very recently, a photoinduced polarization suppression 
was observed in croconic acid and it was regarded as a light-induced polarization inversion of protons \cite{iwano}. 
However, at present, we are not aware of any experimental achievement regarding 
a light-induced electronic ferroelectric inversion. 
\par
(TMTTF)$_{2}X$F$_{6}$ (TMTTF = (tetramethyltetrathiafulvalene), $X$=P, As, Sb, Ta) is known as one of the quasi-1D 
quarter-filled organic conductors and it has rich physical phases \cite{Gex1,Gex2,Gex3,Gex4,Gex5,Gth1,Gth2,Gex6,Ogata}. 
In particular, the bulk electronic ferroelectricity of (TMTTF)$_{2}$PF$_{6}$ caused by finite charge disproportion $\delta_{\rm co}$ 
has been experimentally depicted \cite{Ferro0,Ferro1,Ferro2,Ferro3,Ferro4} in both a CO phase and spin-Peierls (SP) phase. 
According to Ref. \cite{Gex2}, the CO and SP phase of (TMTTF)$_{2}$PF$_{6}$ have been achieved below 67 K and 19 K, respectively.  
In our study, representing $\rho_{\rm rich}$ ($\rho_{\rm poor}$) as the rich (poor) value of the charge 
of the two closest TMTTF molecules forming a dimer, 
$\delta_{\rm co}\equiv\rho_{\rm rich}-\rho_{\rm poor}\geq 0$ is treated. 
The observed finite $\delta_{\rm co}$ values in (TMTTF)$_{2}$PF$_{6}$ below 67 K \cite{Sawa,Tetr,SP2} indicate that 
the SP phase also has the characteristics of the CO phase of (TMTTF)$_{2}$PF$_{6}$. 
In the theoretical works on other materials in terms of the PIPTs \cite{Nex,ThDM,FE2,FE4}, 
the photoexcitations associated with the collective excitations of charges, namely, multi-excitons, have been discussed. 
When these concepts are applied to (TMTTF)$_{2}$PF$_{6}$, photoexcitations of the polarization (P)-inverted domains 
from its ferroelectric ground state are strongly expected, which will lead to macroscopic polarization inversion, 
as shown in Figs. \ref{fig1}(a)--(c). 
Here, note that a dimer corresponds to a unit cell. Defining the total number of dimers as $D$, 
the bulk ferroelectric inversion is achieved when charges with $D\delta_{\rm co}$ move from the CO ground state 
in the entire system. 
\par
To realize such macroscopic polarization inversion in (TMTTF)$_{2}$PF$_{6}$, the most important issue is to know accurately 
the nature of the low-energy optical excitations. For this purpose, several optical conductivities have already 
been observed \cite{ODexp,ODexp2,Iwai}. However, little is known about the pure electronic excitations related to the peaks of 
the optical spectra. In this regard, we first observed the optical conductivity of (TMTTF)$_{2}$PF$_{6}$ at 10 K (in the SP phase) 
and estimated $\omega_{\rm CT}\sim 0.128$ eV as the pure electronic excitation energy, as shown in 
Fig. \ref{fig1}(d). Details of this measurement are explained in Appendix \ref{AppdxA}. A single crystal of (TMTTF)$_{2}$PF$_{6}$ 
was prepared by a previous method \cite{nkmr1,nkmr2}. The complete structure of our spectrum, as shown as a solid line in 
Fig. \ref{fig1}(d), 
is very similar to the previous spectra of (TMTTF)$_{2}$PF$_{6}$ at 20 K (in the CO phase) \cite{ODexp,ODexp2}. This suggests that 
the pure electronic photoexcited state from the CO ground state can be physically considered as almost the same as that 
from the SP ground state. 
In the following sections of this article, we introduce our theoretical analyses, particularly of the optical conductivity spectrum 
in (TMTTF)$_{2}$PF$_{6}$, and discuss the nature of the observed peak structure exhibited as a chained line in 
Fig. \ref{fig1}(d). 
Throughout this paper, we consider $\hbar=e=1$ and lattice constant $=1$ for simplicity. 
%
%
%
\section{Formulation}
\par
Now, we consider a dimerized 1D chain model with even $N_{s}$ sites, which is a quarter-filled hole
system. An equal population of spins ($N_{\uparrow}=N_{\downarrow}=N_{s}/4$) is assumed at absolute zero temperature. 
Using model-specified parameters $V_{\rm eff}$ and $V_{\rm edge}$, our Hamiltonian $H$ is written as follows: 
\begin{align}
H &\equiv H_{t} + H_{\rm Coulomb} + V_{\rm eff}\sum_{j:{\rm even}}n_{j}+V_{\rm edge}n_{N_{s}}, 
\label{eqs1} \\
H_{t} &= -\sum_{j,\sigma}t_{j}\left[
c_{j+1,\sigma}^{\dagger}c_{j,\sigma} + c_{j,\sigma}^{\dagger}c_{j+1,\sigma}
\right] \label{S1}, \\ 
H_{\rm Coulomb} &= U\sum_{j}n_{j,\uparrow}n_{j,\downarrow} + V\sum_{j}n_{j+1}n_{j}, \label{S2}
\end{align} 
where $c_{j,\sigma}^{(\dagger)}$ denotes the annihilation (creation) operator of a hole with spin $\sigma=\uparrow, \downarrow$ 
at the $j$-th site and $n_{j}\equiv n_{j,\uparrow} + n_{j,\downarrow}$ represents the $j$-th site density operator 
($n_{j,\sigma}\equiv c_{j,\sigma}^{\dagger}c_{j,\sigma}$). $j$ denotes a highest occupied molecular orbital (HOMO) of a 
TMTTF molecule. Because each dimer has three electrons in HOMOs and the band consists of HOMOs, the system is regarded 
as a (third) quarter-filling in terms of holes (electrons). The dimerization of the system is treated 
in term $t_{j}$ where $t_{j}\equiv t_{1}$ ($t_{2}$) for even (odd) $j$ represents an inter (intra)-dimer transfer integral. 
From a density functional theory (DFT) calculation of (TMTTF)$_{2}$PF$_{6}$ at 4 K \cite{DFTcal}, we select $t_{1}=0.1686$ eV 
and $t_{2}=0.1912$ eV. Referring to the reported Coulomb repulsive interaction strengths for 
(TMTTF)$_{2}X$-type compounds \cite{Suzumura,Ogata}, we basically use $U=1$ eV and $V=0.2$--0.6 eV. 
\par
Within the framework of the linear response theory, optical conductivity with respect to photon energy $\omega>0$ 
and infinitesimally small positive number $\eta$ is written as 
\begin{equation}
\sigma_{1}(\omega) = -\frac{1}{N_{s}\omega}{\rm Im}\left[
\langle\psi_{0}|J\frac{1}{\omega+i\eta+E_{0}-H}J|\psi_{0}\rangle
\right], 
\end{equation}
where 
\begin{equation}
J\equiv i\sum_{j,\sigma}t_{j}[
c_{j+1,\sigma}^{\dagger}c_{j,\sigma} - c_{j,\sigma}^{\dagger}c_{j+1,\sigma}]
\label{eqs5}
\end{equation}
denotes a charge--current operator, $E_{0}$ represents the ground-state energy, and $|\psi_{0}\rangle$ is 
the ground-state wavefunction. For computational problems, $\eta/t_{2}=0.05$ ($\sim 0.01$ eV) is used. 
\par 
$\sigma_{1}(\omega)$ is computed by the dynamical density-matrix renormalization group (dynamical DMRG or DDMRG) 
scheme \cite{DDMRGJ} under the open boundary condition (OBC). 
In general, although the numerical accuracy of a DMRG \cite{DMRG} calculation under 
the OBC is better than that under the periodic boundary condition (PBC), the charges around the edges under the OBC are rich 
because of breaking of the translational symmetry of the system. 
Although several approaches have been proposed to avoid this unphysical problem to some extent \cite{OBC1,OBC2,OBC3}, 
in this study, we apply potential $V_{\rm edge}$ at the edge site \cite{OBCED} as one of its solutions and 
fix $V_{\rm edge}=50t_{2}$. The value of $V_{\rm edge}=50t_{2}$ is chosen as small as possible to satisfy the condition 
that $E_{0}$ of all the calculations hardly depend on $V_{\rm edge}$ due to unpermitted $V_{\rm edge}\rightarrow +\infty$. 
Because the charge at the $N_{s}$-th site is poor at the $V_{\rm edge}$, the CO ground state considered here 
has a charge-rich (poor) site at the first ($N_{s}$-th) site. 
\par 
Our calculations are done with $N_{s}=100$ (50 dimers). This value is enough large to satisfy with $N_{s}+1\sim N_{s}$ 
(the system size under the OBC) and to quantitatively estimate the bulk properties although 
finite size effects still remain in the order of $1/N_{s}$. The truncation number of density matrices is 400 in our all the DMRG 
and DDMRG calculations. All the sweep processes stopped when the numerical relative error of adjacent sweeps was 
less than 10$^{-6}$ for $E_{0}$ and 10$^{-3}$ for $\sigma_{1}(\omega)$. 
\par
We introduce number of photoexcited charges $N_{\rm ex}$ \cite{Nex} to discuss the relationship between a photoexcited state 
and the collective excitations of the charges. Using 
\begin{equation}
|\psi(\omega)\rangle \equiv \frac{1}{\mathcal{N}}\frac{\eta}{(\omega+E_{0}-H)^{2}+\eta^{2}}J|\psi_{0}\rangle, 
\end{equation}
where $\mathcal{N}$ denotes a normalization factor of $|\psi(\omega)\rangle$, 
\begin{equation}
N_{\rm ex} \equiv \sum_{j:{\rm even}}[\langle\psi(\omega)|n_{j}|\psi(\omega)\rangle-\langle\psi_{0}|n_{j}|\psi_{0}\rangle]
\end{equation}
can be defined. Here, $\langle\phi|n_{j}|\phi\rangle \; (\phi=\psi_{0}, \psi(\omega))$ corresponds to the site density at the $j$-th site. 
Because we consider weak photoexcitations and a single photon injected into the system, $N_{\rm ex}>1$ denotes 
the occurrence of collective excitation. We also theoretically estimate the charge disproportion by 
\begin{equation}
\delta_{\rm co}\equiv \frac{1}{2}\sum_{j=49,51}|\langle\phi|n_{j}-n_{j+1}|\phi\rangle| \quad (0\leq \delta_{\rm co}\leq 1). 
\end{equation}
Because the center of the system gives most accurate expectation values of localized operators by DMRG calculations 
under the OBC, we choose the system centered two dimers for calculating $\delta_{\rm co}$. 
%
%
\section{Results and Discussions for $V_{\rm eff}=0$}
%
%
\par
\begin{figure}[t]
\begin{center}
\includegraphics[width=10cm,keepaspectratio]{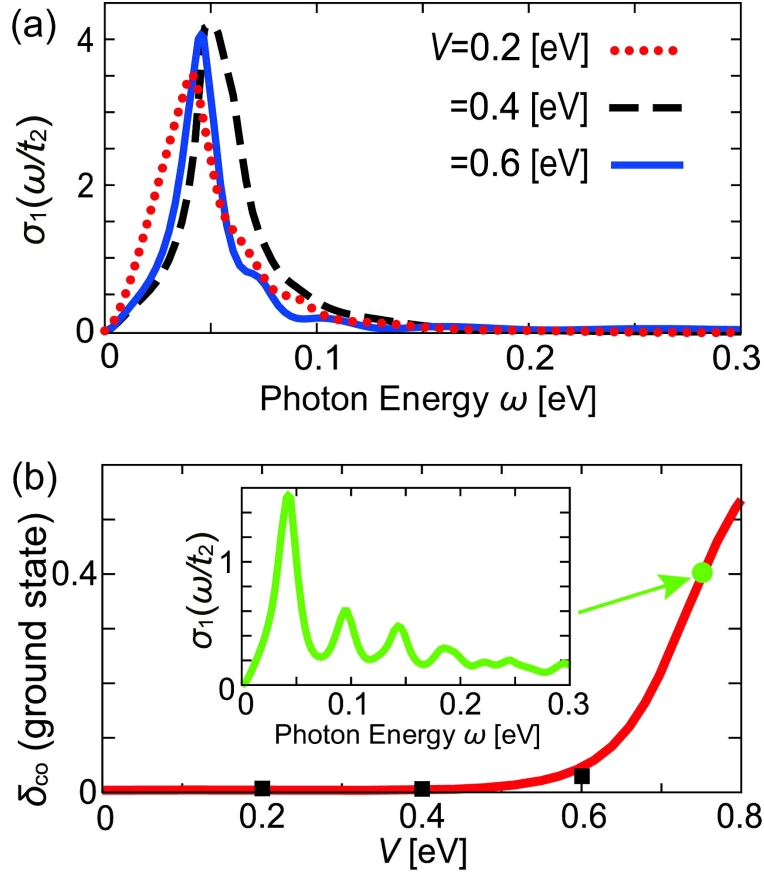}
\caption{DMRG and DDMRG calculations for $V_{\rm eff}=0$ at $N_{s}=100$. 
(a) Results for $\sigma_{1}(\omega/t_{2})$. 
(b) $\delta_{\rm co}$ of ground states. The parameter set of $t_{1}=0.20$ eV, $t_{2}=0.22$ eV, and $U=2.2$ eV 
\cite{DFTcal2} is used only for calculations presented in this figure. $\delta_{\rm co}$ values at the ground states for 
the parameter sets in (a) are plotted as filled squares for comparison. The inset is $\sigma_{1}(\omega/t_{2})$ with 
$t_{1}=0.20$ eV, $t_{2}=0.22$ eV, $U=2.2$ eV \cite{DFTcal2}, and 
$V=0.75$ eV for giving $\delta_{\rm co}=0.40$ in the ground state. 
}
\label{fig2}
\end{center}
\end{figure}
\par
\par
We first show the theoretical results for $\sigma_{1}(\omega)$ using several realistic values for $V$ in the case of $V_{\rm eff}=0$ 
(the conventional model) as shown in 
Fig. \ref{fig2}(a). 
As it can be seen, a single peak of $\sigma_{1}(\omega)$ appears around the so-called dimerization gap of 
$\omega_{\rm d}\equiv 2|t_{1}-t_{2}|\sim 0.045$ eV, which corresponds to the minimum gap of free dispersions. Although this is 
supported by another DDMRG calculation \cite{DDMRGJ2} with a different parameter set \cite{DDMRGJ3,DDMRGJ4}, 
$\omega_{\rm d}$ deviates from $\omega_{\rm CT}$. 
In addition, in the ground state, $\delta_{\rm co}\sim 0.03$ is the maximum value in our calculation and does not reproduce 
$\delta_{\rm co}=0.40$, which was recently observed in an X-ray diffraction experiment at 30 K \cite{Sawa}. 
To reproduce $\delta_{\rm co}=0.40$, we recalculate $\delta_{\rm co}$ as a function of $V$ by utilizing a different parameter set 
estimated by another DFT calculation \cite{DFTcal2}, namely, $U=2.2$ eV, $t_1=0.20$ eV, and $t_2=0.22$ eV. 
The results are shown in Fig. \ref{fig2}(b), 
and we determine the best parameter of $V=0.75$ eV. However, the complete structure of $\sigma_{1}(\omega)$ at $V=0.75$ eV 
shown in the inset of 
Fig. \ref{fig2}(b) 
deviates from our observation at 10 K, as shown in 
Fig. \ref{fig1}(d). 
In particular, the broad spectral shape significantly differs from the observed single peak, and we
interpret the former feature as the exaggerated collectiveness of the excitations, which will be discussed subsequently. Thus, 
the conventional model ($V_{\rm eff}=0$) should be modified to some extent. 
%
%
\section{Results and Discussions for $V_{\rm eff}\neq 0$}
%
%
\par
\begin{figure}[t]
\begin{center}
\includegraphics[width=10cm,keepaspectratio]{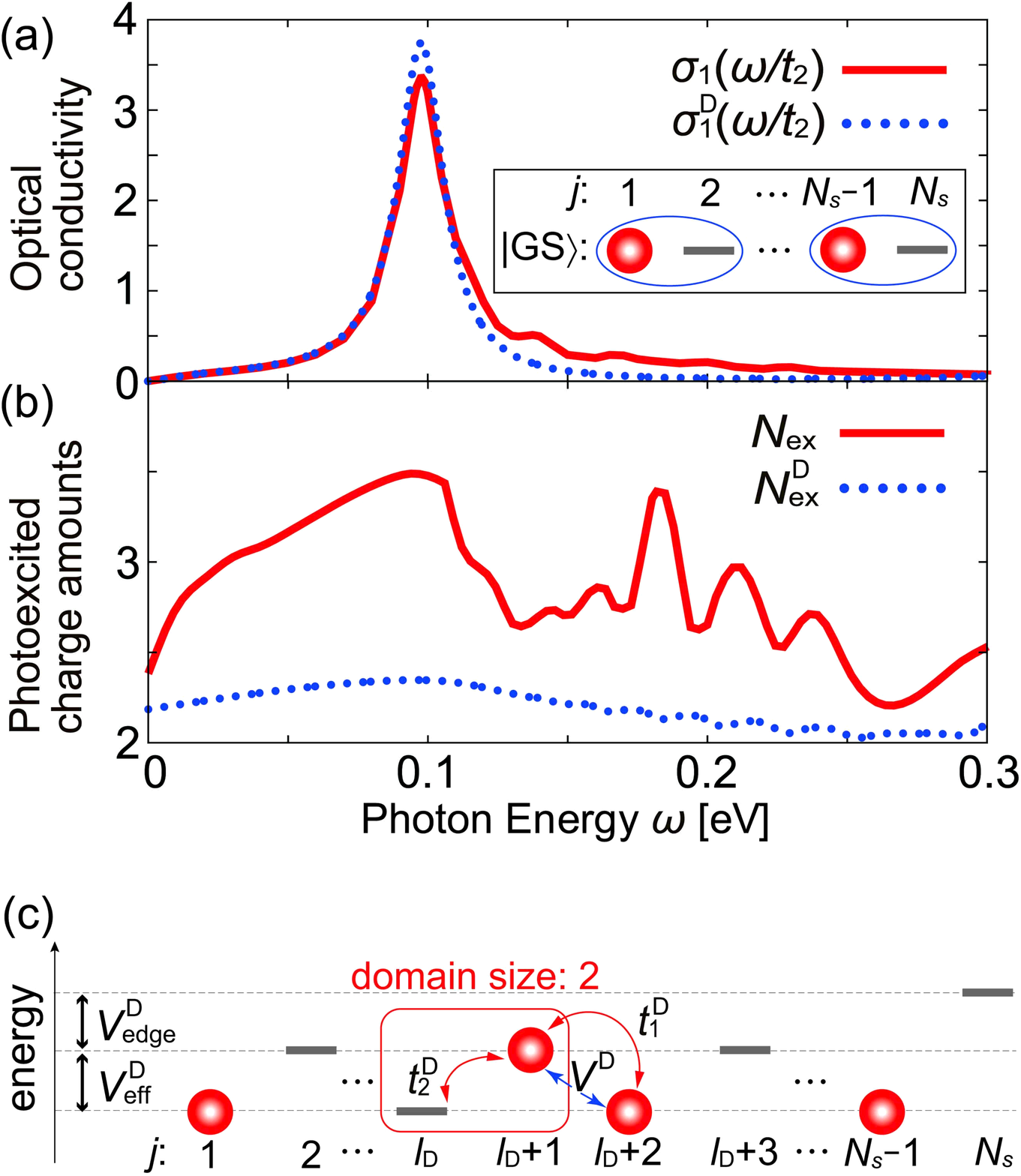}
\caption{Calculations of $V_{\rm eff}\neq 0$ at $N_{s}=100$. 
(a), (b) DDMRG results of $\sigma_{1}(\omega), N_{\rm ex}$ (solid lines) and 
$\sigma_{1}^{\rm D}(\omega), N_{\rm ex}^{\rm D}$ by using our effective model (dotted lines). 
The inset of (a) is a schematic of $|{\rm GS}\rangle$ in our effective model ($\delta_{\rm co}=1$). 
The circles and horizontal bars represent single charges and empty sites, respectively. 
(c) Schematic energy diagram of basis $|l_{\rm D},n\rangle$ with $n=1$ and odd $l_{\rm D}$. Only $|l_{\rm D}:{\rm odd},n\rangle$ 
states are generated from intra-dimer hopping. 
}
\label{fig3}
\end{center}
\end{figure}
\par
\par
As an alternative approach for reproducing $\delta_{\rm co}=0.40$ \cite{Sawa}, we introduce the $V_{\rm eff}$ term, 
which increases $\delta_{\rm co}$. 
Next, we employ $V/t_{2}=3.138$ ($V=0.6$ eV) because it gives the maximum value, $\delta_{\rm co}\sim 0.03$, 
in the ground state with $V_{\rm eff}=0$. We find $V_{\rm eff}/t_{2}=0.086$ as the best value. 
The results of $\sigma_{1}(\omega)$ and $N_{\rm ex}$ with $V/t_{2}=3.138$, $V_{\rm eff}/t_{2}=0.086$ are shown in 
Figs. \ref{fig3}(a) and (b), respectively. 
A sharp peak structure of $\sigma_{1}(\omega)$ can be seen to arise around $0.10$ eV, and this is clearly closer to 
$\omega_{\rm CT}$ than $\omega_{\rm d}$. 
Furthermore, in 
Fig. \ref{fig3}(b), 
$N_{\rm ex}\geq 2$ denotes that all the photoexcited states are the collective excitations of the charges and 
that maximum value $N_{\rm ex}\sim 3.5$ appears at the sharp peak of $\sigma_{1}(\omega)$. As mentioned above, because 
$N_{\rm ex} = \delta_{\rm co}N_{s}/2 = 20$ corresponds to the bulk ferroelectric inversion, a polarization inversion over 
$3.5/\delta_{\rm co}\sim 9$ unit cells ($3.5/20\sim 18\%$ of all the charges) can be achieved at the peak. 
To understand this collective excitation at the peak, we compare the site density of this peak state with that of the ground state in 
Fig. \ref{fig4}(a), 
and we find that $\delta_{\rm co}$ decreases to approximately $0.13$. This reduction in $\delta_{\rm co}$ at the peak 
can be explained by two scenarios as follows. 
The photoexcited state at the peak partially includes a P-inverted domain ($\delta_{\rm co}\rightarrow -\delta_{\rm co}\neq 0$) 
or dimer--Mott (DM)-insulating state ($\delta_{\rm co}=0$). In this article, we discuss only the former scenario by extending 
the effective model proposed in Refs. \cite{Nex,ThDM}. 
However, the above case of a DM-insulating state is insignificant, as discussed in Appendix \ref{AppdxC}. 
In addition, we also discuss EIMV coupling \cite{eph1,eph2,eph3} as one of the origins of unconventional term 
$V_{\rm eff}$ in Appendix \ref{AppdxB}. $V_{\rm eff}$ under EIMV coupling is physically related to an effective potential 
representing the deformed molecular orbitals with $\delta_{\rm co}\neq 0$. 
%
%
\section{Effective model analysis for $V_{\rm eff}\neq 0$}
%
%
\par
\begin{figure}[t]
\begin{center}
\includegraphics[width=10cm,keepaspectratio]{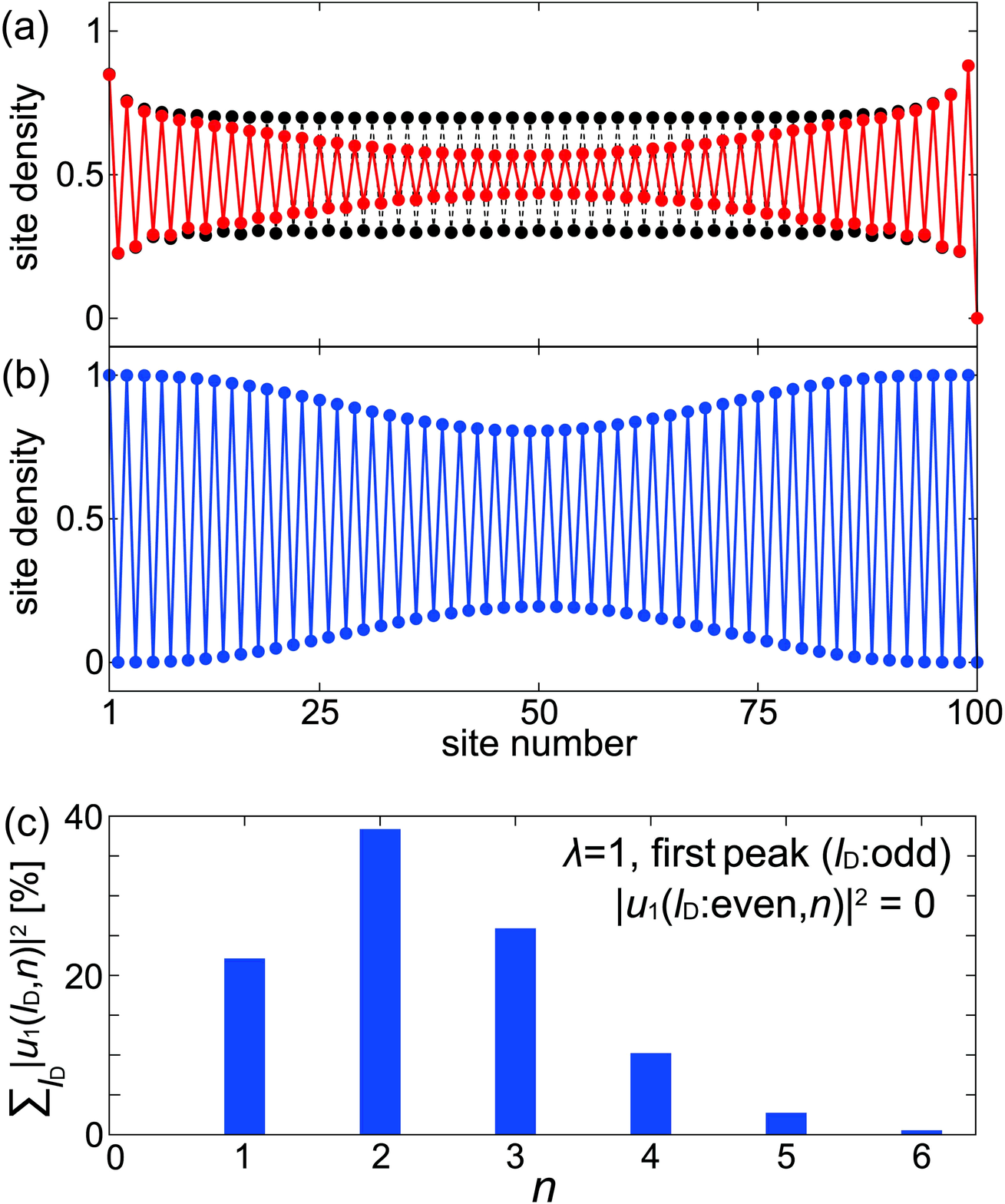}
\caption{Properties of the photoexcited states at the peak of both 
$\sigma_{1}(\omega)$ and $\sigma_{1}^{\rm D}(\omega)$ with $V_{\rm eff}\neq 0$ at $N_{s}=100$. 
(a) Site densities of the ground state (dotted line) and peak state (solid line) by the DDMRG method.  
(b) Site density of the peak state by our effective model. 
(c) Probability weights of the peak state by our effective model with respect to basis $|l_{\rm D},n\rangle$ 
for odd $l_{\rm D}$. Probability weights for even $l_{\rm D}$ are absent. 
}
\label{fig4}
\end{center}
\end{figure}
\par
\par
Our effective model under the OBC, assumes the CO ground state obtained by the DMRG method ($\delta_{\rm co}=0.40$) 
as the charge localized limit, $\delta_{\rm co}=1$. The normalized ground-state wavefunction of this model, $|{\rm GS}\rangle$, 
only contains the charges at the odd sites. Using the site density operator at the $j$-th site, $n_{j}^{\rm D}$, 
$\langle{\rm GS}|n_{j}^{\rm D}|{\rm GS}\rangle=1$ $(0)$ for odd (even) $j$ is satisfied as approximately sketched 
in the inset of Fig. \ref{fig3}(a). All the physical parameters also differ from those of $H$, and in particular, 
$U$ vanishes in this model. When we define basis $|l_{\rm D},n\rangle$ as the photoexcited state with 
a single P-inverted domain continuously arranged in $2n$ sites with starting site $l_{\rm D}$, 
the Hamiltonian of our effective model is described as  
\begin{equation}
H_{\rm dmn} \equiv - \sum_{l_{\rm D},n}t(l_{\rm D}) [|l_{\rm D}-2,n+1\rangle\langle l_{\rm D},n| 
+ |l_{\rm D},n+1\rangle\langle l_{\rm D},n| + h. c.] 
+ \sum_{l_{\rm D},n}E(n)|l_{\rm D},n\rangle\langle l_{\rm D},n|, 
\end{equation}
where $t(l_{\rm D})\equiv t_{1}^{\rm D}$ ($t_{2}^{\rm D}$) for even (odd) $l_{\rm D}$ and 
\begin{equation}
E(n) = \begin{cases}
         nV_{\rm eff}^{\rm D} + V_{\rm  edge}^{\rm D} & (l_{\rm D}=N_{s}-2n+1) \\
         V^{\rm D} + nV_{\rm eff}^{\rm D} & (\rm{otherwise}). 
  \end{cases}
\end{equation}
Here, $t_{1}^{\rm D}$ ($t_{2}^{\rm D}$) denotes an inter (intra)-dimer transfer integral. A schematic is shown 
in Fig. \ref{fig3}(c). 
Eigenenergies $\varepsilon_{\lambda}$ and eigenstates $|\lambda\rangle$ ($1\leq\lambda\leq (N_{s}/2)^{2}$) satisfy 
$H_{\rm dmn}|\lambda\rangle=\varepsilon_{\lambda}|\lambda\rangle\equiv\sum_{l_{\rm D},n}u_{\lambda}(l_{\rm D},n)|l_{\rm D},n\rangle$. 
Introducing the charge--current operator of this model $J^{\rm D}$ and 
$|\psi_{1}\rangle\equiv J^{\rm D}|{\rm GS}\rangle=i\sum_{l_{\rm D}}(-1)^{l_{\rm D}-1}t(l_{\rm D})|l_{\rm D},1\rangle$, 
the optical conductivity of this model is defined as 
\begin{equation}
\sigma_{1}^{\rm D}(\omega) = \frac{\eta}{N_{s}\omega}\sum_{\lambda}
\frac{|\langle\lambda|\psi_{1}\rangle|^{2} }{(\omega-\varepsilon_{\lambda})^{2}+\eta^{2}}
\equiv \frac{\langle\psi_{1}|\omega\rangle}{\mathcal{C}N_{s}\omega}, 
\end{equation}
where $\mathcal{C}$ is determined by $1=\langle\omega|\omega\rangle$ and number of photoexcited charges 
\begin{equation}
N_{\rm ex}^{\rm D} \equiv \sum_{j:{\rm even}}[\langle\omega|n_{j}^{\rm D}|\omega\rangle-
\langle{\rm GS}|n_{j}^{\rm D}|{\rm GS}\rangle] = \sum_{j:{\rm even}}\langle\omega|n_{j}^{\rm D}|\omega\rangle
\end{equation}
can be also defined. 
\par
To relate the nature of the photoexcited state at the peak of $\sigma_{1}(\omega)$ 
to that of $\sigma_{1}^{\rm D}(\omega)$, $\sigma_{1}^{\rm D}(\omega)$ and $N_{\rm ex}^{\rm D}$ should resemble 
$\sigma_{1}(\omega)$ and $N_{\rm ex}$, respectively, most accurately. 
Employing $N_{s}=100$, $t_{2}^{\rm D}=t_{2}=0.1912$ eV, $V_{\rm edge}^{\rm D}/t_{2}^{\rm D}=50$, and 
$\eta/t_{2}^{\rm D}=0.05$, we succeeded in reproducing $\sigma_{1}(\omega)$ of $V_{\rm eff}\neq 0$ with 
$t_{1}^{\rm D}/t_{2}^{\rm D}=0.600$, $V^{\rm D}/t_{2}^{\rm D}=2.615$, and $V_{\rm eff}^{\rm D}/t_{2}^{\rm D}=0.528$ as shown in 
Fig. \ref{fig3}(a). 
Although $N_{\rm ex}^{\rm D}$ underestimates $N_{\rm ex}$ with $V_{\rm eff}\neq 0$ as illustrated in 
Fig. \ref{fig3}(b), 
the overall behavior of $N_{\rm ex}^{\rm D}$ is qualitatively consistent with that of $N_{\rm ex}$ in terms of the collective excitation 
of the charges ($N_{\rm ex}^{\rm D}>2$) in the entire $\omega$ region. $N_{\rm ex}^{\rm D}\sim 2.3$ is the maximum value and 
appears at the peak of $\sigma_{1}^{\rm D}(\omega)$. 
In addition, although the site density at the peak as displayed in Fig. \ref{fig4}(b) 
differs from that obtained by the DDMRG scheme in Fig. \ref{fig4}(a) 
owing to our assumption of $\delta_{\rm co}=1$ at the ground state, the dip structure located around the center of the system 
is consistent with the DDMRG result. According to the above results, the peak state of $\sigma_{1}(\omega)$ can be regarded 
as that of $\sigma_{1}^{\rm D}(\omega)$, which corresponds to $\lambda=1$ eigenstate with 
$\sum_{l_{\rm D}={\rm odd}}\sum_{1\leq n\leq 6}|u_{1}(l_{\rm D}={\rm odd},n)|^{2}\sim 99.9\%$ and 
$u_{1}(l_{\rm D}={\rm even},n)=0$, as shown in 
Fig. \ref{fig4}(c). 
This implies that the photoexcited state is generated by an intra-dimer hopping and consists of 
the superposition of the P-inverted domains over 1--6 unit cells. 
\par
Regarding the case of $U=2.2$ eV mentioned already, we find much larger $N_{\rm ex}$ values at the peaks of 
the corresponding spectrum 
(inset of Fig. \ref{fig2}(b)), 
which are in the range 6--9. 
Because these large $N_{\rm ex}$ values indicate a strong collectiveness of the charge excitations \cite{Nex}, we determined that 
the present system lies in the category of modest collectiveness.
%
%
\section{Conclusions}
\par
To conclude, we have investigated a photoexcited state from the CO ground state of (TMTTF)$_{2}$PF$_{6}$. We found that 
the calculated spectrum based on a quarter-filled 1D effective model ($V_{\rm eff}\neq 0$) reproduce the experimental spectrum 
of the CO ground state. We clarified that the electronic component of the optical conductivity had a single significant peak 
around 0.10 eV and that the photoexcited state at the peak could be regarded as a superimposed state of the P-inverted 
domains with a modest collectiveness. 
\par
For the photoexcited state at the peak, approximately 18$\%$ of the charges in the system contribute to the P-inverted 
domains generated by a single photon (weak photoexcitation). Regarding a relative change of electric polarization ($\Delta P/P$) 
related to measurements of a second-harmonic generation, the photoexcitation corresponds to $\Delta P/P\sim 36\%$. 
The value of $36\%$ is clearly much larger than $\Delta P/P\sim 2$--$10\%$ of other ferroelectric organic materials for a weak 
photoexcitation \cite{iwano,FEex1}. Moreover, multi-photons (strong photoexcitation) can enhance $\Delta P/P$ and possibly 
generate a P-inverted domain spreading over the entire system, which is simply the achievement of bulk ferroelectric inversion. 
Therefore, examining strongly photoexcited effects is one of the crucial and challenging future tasks. 
However, our results adequately showed that (TMTTF)$_{2}$PF$_{6}$ could be one of promising materials for 
applications in optical switching devices and memories in the context of such macroscopic manipulation of ferroelectricity. 
%
%
\begin{acknowledgments}
This work was supported by JST CREST in Japan (Grant No. JPMJCR1661). 
K.I. was supported by the Grant-in-Aid for Scientific Research from JSPS in Japan (Grant No. JP17K05509). 
The computations were partially performed at the Research Center for Computational Science, Okazaki, Japan. 
\end{acknowledgments}
%
%
\appendix
\renewcommand{\thetable}{\Alph{section}\arabic{table}}
\renewcommand{\thefigure}{\Alph{section}\arabic{figure}}
\section{Details of our experiment}
\label{AppdxA}
\setcounter{figure}{0}
\par
\begin{figure}[t]
\centering
\includegraphics[width=12cm,keepaspectratio]{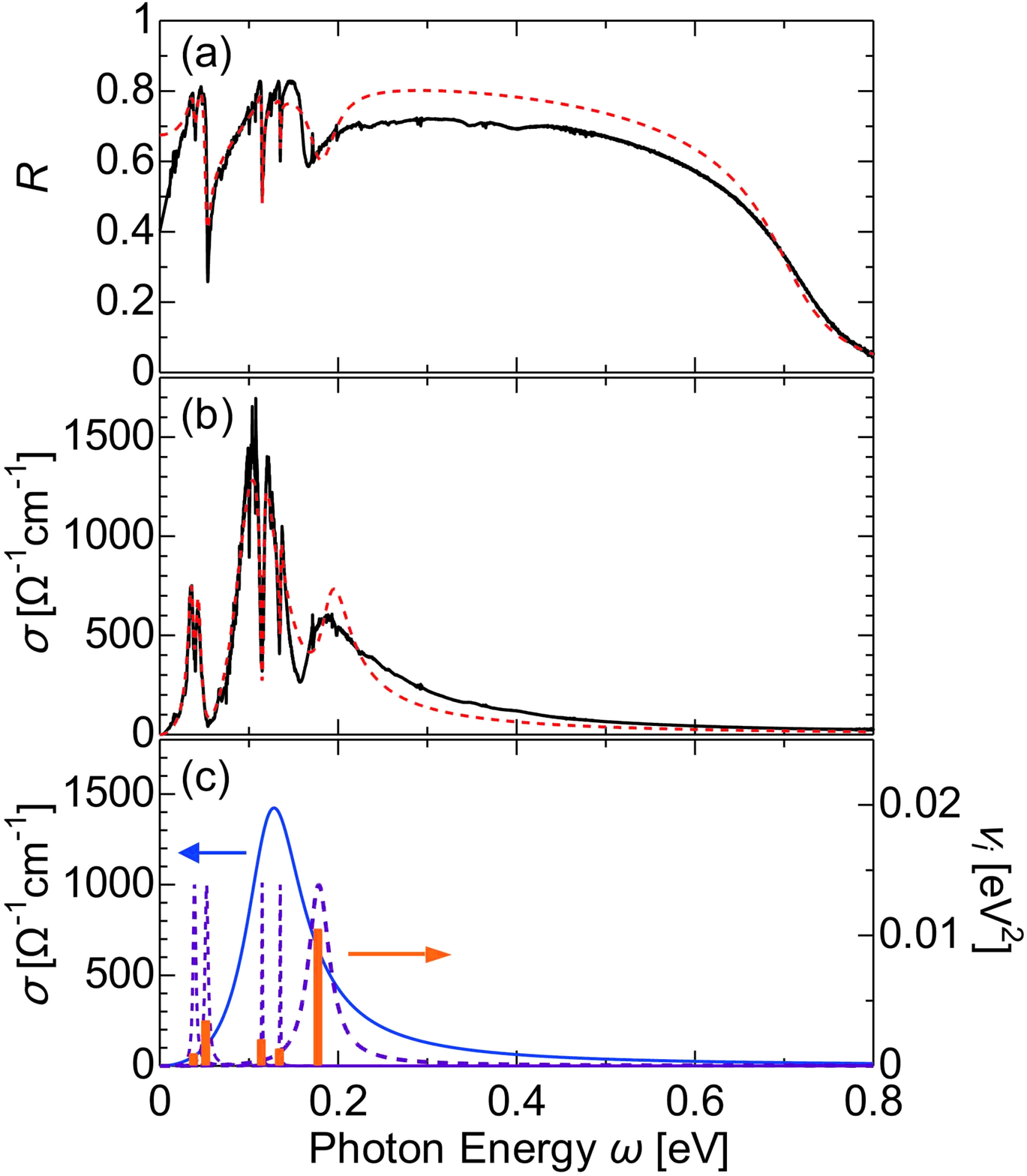}
\caption{(a) Polarized reflectivity and (b) optical conductivity spectrum for $E\parallel a$ at 10 K (the solid lines). 
The dashed lines show the fitting curves. (c) Calculated spectra of the CT transition (the solid line) and phonons 
(the dashed lines). The magnitude of the latter is normalized at 1000 $\Omega^{-1}$cm$^{-1}$. The coupling between each phonon 
and the CT transition is shown as a bar. 
}
\label{sfig1}
\end{figure}
\par
\par
\begin{figure}[t]
\begin{center}
\includegraphics[width=8cm,keepaspectratio]{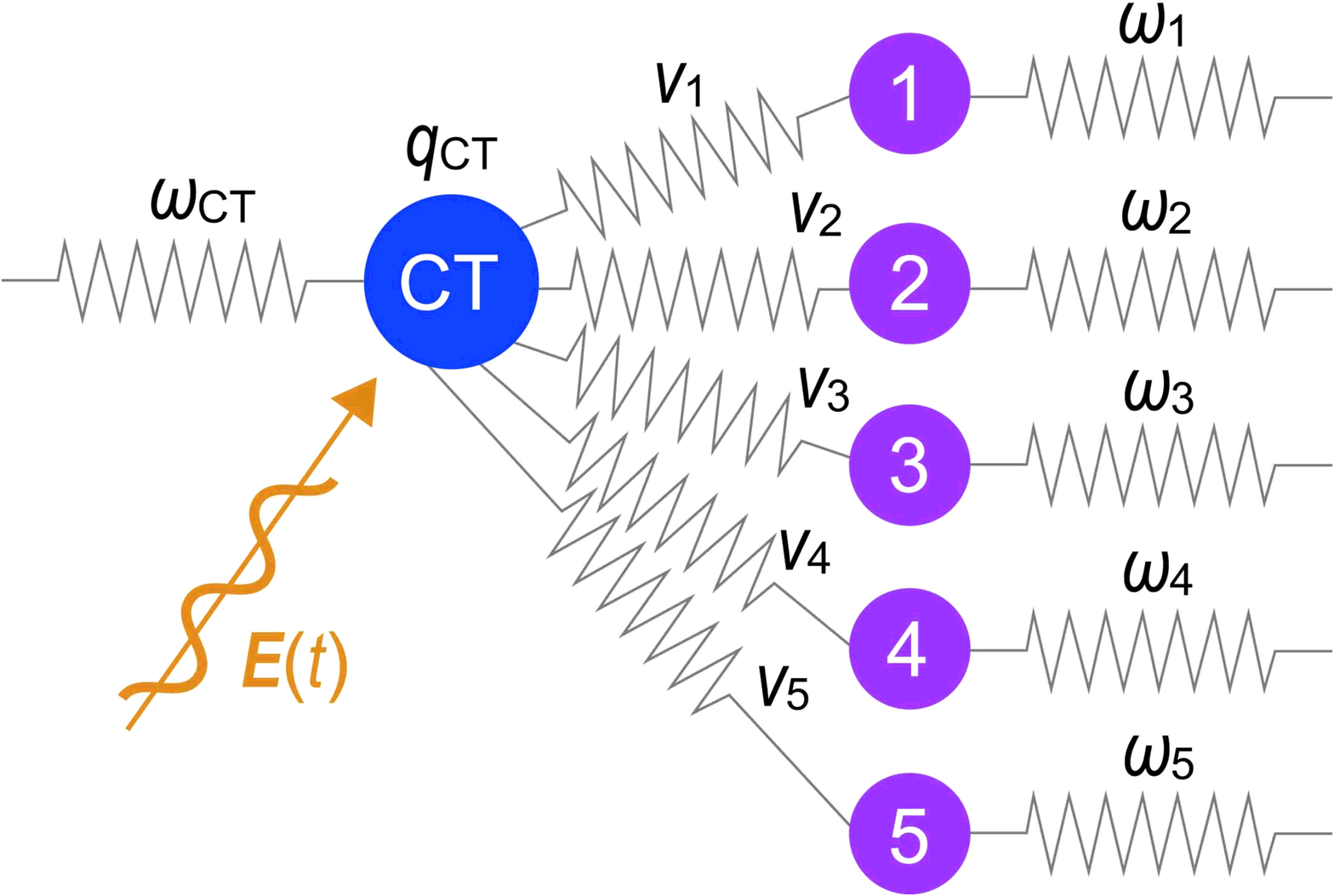}
\caption{Schematic of the Fano interference originating from the EIMV coupling. 
}
\label{sfig2}
\end{center}
\end{figure}
\par
\par
To evaluate the pure electronic excitation energy, we perform a fitting analysis of the optical conductivity spectrum, 
$\sigma(\omega)$, as shown in Fig. \ref{fig1}(d). It is well known that the reflectivity spectrum of a 
TMTTF (tetramethyltetrathiafulvalene) salt has a complex structure in the lower energy region, which is attributed to 
the Fano interference originating from 
the electron-intramolecular vibration (EIMV) coupling between the charge transfer (CT) transition and Raman active 
intramolecular vibration modes below 0.2 eV \cite{Iwai}. Fitting analyses based on the dimer model considering this effect 
have been performed \cite{Iwai,dmm1,dmm2}. In this section, we expand this method to analyze the optical spectrum. 
\par
$R$ in Fig. \ref{sfig1}(a) shows the reflectivity spectrum of (TMTTF)$_{2}$PF$_{6}$ for the electric field polarized parallel to 
the $a$-axis ($E\parallel a$), which is measured by a Fourier transform infrared spectrometer. $\sigma(\omega)$ is obtained 
by the Kramers--Kronig transformation of the reflectivity spectrum as shown in 
Fig. \ref{sfig1}(b). 
Considering the EIMV coupling effect in the framework of the Fano interference, we perform a fitting analysis of 
these spectra. The Fano interference is known to be analogous to the toy model considering classical harmonic oscillators 
interacting each other \cite{dmm3}. In this model, the vibration modes of (TMTTF)$_{2}$PF$_{6}$ can be described as Fig. \ref{sfig2}. 
The purely electronic CT transition without the EIMV coupling is regarded as an oscillator with charge $q_{\rm CT}$ and 
eigenfrequency $\omega_{\rm CT}$. In addition to this CT oscillator, infrared inactive intramolecular vibrations are introduced 
as oscillators $j$ with eigenfrequencies $\omega_{j} \; (j=1, 2, 3, 4, 5)$ without charges. Oscillator $j$ is coupled with 
the CT oscillator via coupling constant $\nu_{j}$. When a light having the electric field $E(t)$ is irradiated, 
only the CT oscillator is directly driven. Subsequently, the coupled vibration of oscillator $j$ is generated 
by the vibration of the CT oscillator via the EIMV coupling. This can be attributed to the infrared activation of the 
original infrared inactive intramolecular vibration modes due to the interaction with the CT transition. 
\par
The equation of motion for this system with external electric field $E(\omega)$ can be expressed as follows: 
\begin{equation}
\begin{split}
\begin{bmatrix}
L_{\rm CT}^{-1} & -\nu_{1} & -\nu_{2} & -\nu_{3} & -\nu_{4} & -\nu_{5} \\
-\nu_{1} & L_{1}^{-1} & 0 & 0 & 0 & 0 \\
-\nu_{2} & 0 & L_{2}^{-1} & 0 & 0 & 0 \\
-\nu_{3} & 0 & 0 & L_{3}^{-1} & 0 & 0 \\
-\nu_{4} & 0 & 0 & 0 & L_{4}^{-1} & 0 \\
-\nu_{5} & 0 & 0 & 0 & 0 & L_{5}^{-1}
\end{bmatrix}
\begin{bmatrix}
x_{\rm CT} \\
x_{1} \\
x_{2} \\
x_{3} \\
x_{4} \\
x_{5}
\end{bmatrix}
= 
\begin{bmatrix}
q_{\rm CT} \\
0 \\
0 \\
0 \\
0 \\
0
\end{bmatrix}
E(\omega), & \\
L_{\rm CT}^{-1} = \omega_{\rm CT}^{2}-\omega^{2}-i\omega\gamma_{\rm CT}, \; 
L_{j}^{-1} = \omega_{j}^{2}-\omega^{2}-i\omega\gamma_{j}, &
\label{ES1}
\end{split}
\end{equation}
where $x_{\rm CT},\; x_{j}$ denote the displacements of the CT oscillator and oscillators $j$, respectively. From Eq. (\ref{ES1}), 
\begin{equation}
x_{\rm CT} = \frac{q_{\rm CT}L_{\rm CT}}
{1-L_{\rm CT}D}E(\omega) \; \left(D = \sum_{j=1}^{5}\nu_{j}^{2}L_{j}\right)
\label{ES2}
\end{equation}
is derived. Consequently, the dielectric function including fitting parameters can be expressed as 
\begin{equation}
\varepsilon(\omega) = \varepsilon_{\infty} + \frac{\mu_{\rm CT}^{2}L_{\rm CT}}{1-L_{\rm CT}D}. 
\label{ES3}
\end{equation}
Here, $\mu_{\rm CT}$ is a parameter proportional to $q_{\rm CT}$, corresponding to the transition intensity. $\varepsilon_{\infty}$ 
denotes the dielectric function of the background. 
\begin{table}[h]
\caption{Fitting parameters in Eq. (\ref{ES3})}
\label{eTBLS1}
\scalebox{1.2}{
  \begin{tabular}{cccc} \hline
    $\varepsilon_{\infty}$ & $\omega_{\rm CT}$ [eV] & $\gamma_{\rm CT}$ [eV] & $\mu_{\rm CT}$ [eV] \\ \hline 
    1.7 & 0.128 & 7.73$\times 10^{-2}$ & 0.905 \\ \hline 
    & & & \\ \hline
    $j$ & $\omega_{j}$ [eV] & $\gamma_{j}$ [eV] & $\nu_{j}$ [eV$^{2}$] \\ \hline 
    1 & 0.178 & 2.88$\times 10^{-2}$ & 1.04$\times 10^{-2}$ \\
    2 & 0.135 & 9.05$\times 10^{-4}$ & 1.23$\times 10^{-3}$ \\
    3 & 0.115 & 9.07$\times 10^{-4}$ & 1.94$\times 10^{-3}$ \\
    4 & 0.0525 & 4.30$\times 10^{-3}$ & 3.39$\times 10^{-3}$ \\
    5 & 0.0389 & 3.79$\times 10^{-3}$ & 8.67$\times 10^{-4}$ \\ \hline
  \end{tabular}
}
\end{table}
\par
The reflectivity and optical conductivity spectra are calculated by $\varepsilon(\omega)$ in Eq. (\ref{ES3}). 
Measured reflectivity $R$ and $\sigma(\omega)$ 
in Fig. \ref{sfig1} 
are well-reproduced by the fitting curves (the dashed lines). The fitting parameters are listed in Table \ref{eTBLS1}. 
The calculated spectra of the CT transition and phonons are displayed 
in Fig. \ref{sfig1}(c). 
From the fitting analysis, the excitation energy of the pure electronic CT excitation, $\omega_{\rm CT}$, is 
evaluated to be 0.128 eV. The discrepancy between the experimental and calculated spectra in the higher energy region of 
$\sigma(\omega)$ is probably caused by the higher complexity of the spectral shape of the pure electronic CT transition 
than that of the single Lorentz oscillator assumed in this model. This is consistent with the result of our work indicating 
that photoexcited states are collective modes of charges. 
%
%
%
%
\section{Supplemental materials on theories}
\setcounter{figure}{0}
\par
Before discussing the main subject of this section, we newly introduce parts of the model Hamiltonian and physical quantities. 
Here we consider $N_{s}$ sites of a one-dimensional (1D) chain model with a quarter-filled hole system and an equal population 
of spins ($N_{\uparrow}=N_{\downarrow}=N_{s}/4$) at absolute zero temperature again. 
In addition to Eqs. (\ref{eqs1})--(\ref{S2}), we newly define parts of the model Hamiltonian as follows: 
\begin{align}
H_{\rm eff}^{\rm even} &= V_{\rm eff}\sum_{j:{\rm even}}n_{j}, \; H_{\rm eff}^{\rm odd} = V_{\rm eff}\sum_{j:{\rm odd}}n_{j}. \label{S3} 
\end{align} 
In this section, Hamiltonian $H_{t}+H_{\rm Coulomb}+H_{\rm eff}^{\rm even}$ is the same as $H$ in 
Eq. (\ref{eqs1}) 
with $V_{\rm edge}=0$. Using given Hamiltonian $\mathcal{H}$ and the charge--current operator $J$ in Eq. (\ref{eqs5}), 
the reduced optical conductivity of given photon energy $\omega>0$ is written as 
\begin{equation}
\sigma_{\rm 1R}(\omega) = -\frac{1}{N_{s}}{\rm Im}\left[
\langle\psi_{0}|J\frac{1}{\omega+i\eta+E_{0}-\mathcal{H}}J|\psi_{0}\rangle
\right] \label{S5}
\end{equation}
within the framework of the linear response theory for $\eta\rightarrow 0+$. Parameters with $V_{\rm eff}\neq 0$ 
in previous sections (namely, $t_{1}/t_{2}=0.882, U/t_{2}=5.230, V/t_{2}=3.138$, and $\eta/t_{2}=0.05$ 
for $t_{2}=0.1912$ eV \cite{Ogata,DFTcal,Suzumura}) are utilized for all the computations in this section. 
\par
In this section, all the calculations are performed by the exact diagonalization (ED) method under the periodic boundary condition 
(PBC) to avoid the edge effects that typically occur under the open boundary condition (OBC). 
Because the ED calculations are limited to a small system size of the order of $N_{s}\sim 20$ for the computational problem, 
the edge effects significantly affect the calculations and so, should be eliminated. 
In contrast to the density-matrix renormalization group (DMRG) \cite{DMRG} and dynamical DMRG (DDMRG) \cite{DDMRGJ} 
methods in previous sections, the ED method for a fixed system size can easily yield the wavefunctions of arbitrary quantum 
states and allow their comparison owing to the unused renormalized Hamiltonians even if the calculations involve 
different physical parameters. This is the reason why we select the ED method in this section. 
%
%
%
%
\subsection{Estimation of $V_{\rm eff}$}
\label{AppdxB}
\par
\begin{figure}[t]
\begin{center}
\includegraphics[width=10cm,keepaspectratio]{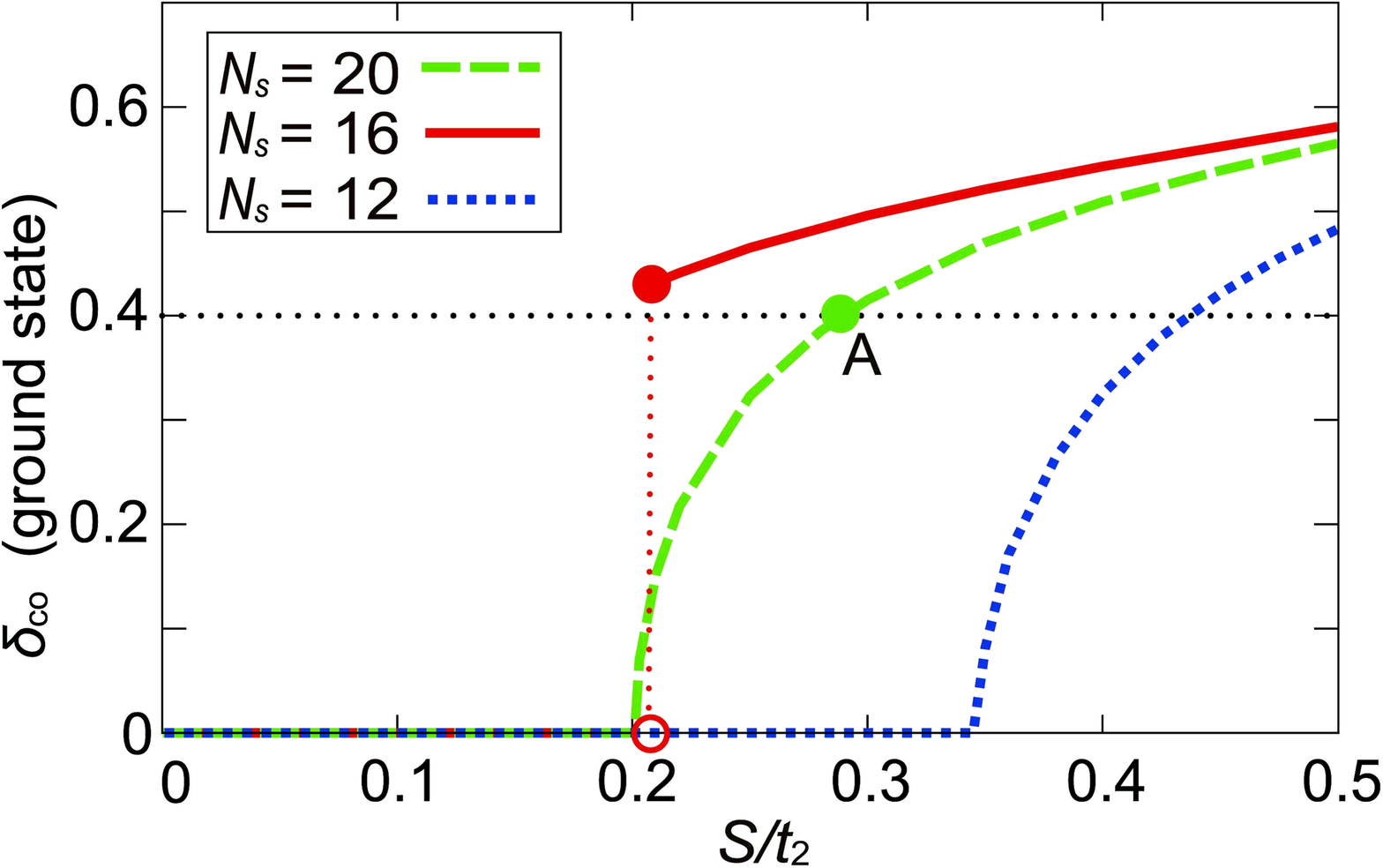}
\caption{Calculations of $\delta_{\rm co}$ for $S/t_{2}$ determined by minimizing $E_{0}^{\rm EIMV}$ in 
Eq. (\ref{S11}) 
for $N_{s}=12, 16, 20$ 
with the ED method under the PBC. The horizontal dotted-line and point A correspond to $\delta_{\rm co}=0.40$ \cite{Sawa}. 
}
\label{sfig3}
\end{center}
\end{figure}
\par
\par 
The aim of this subsection is to discuss one of the origins of $V_{\rm eff}=0.086t_{2}$ in previous sections. 
As mentioned already, an unconventional term, $V_{\rm eff}$, is introduced for reproducing the experimental data 
and indeed consider several origins of $V_{\rm eff}$ such as the quasi-two-dimensional effects from the Coulomb interactions 
between 1D chains. However, we consider that one of the candidates for the origin of $V_{\rm eff}$ is 
the electron-intramolecular vibration (EIMV) coupling, which is one of 
the effective electron--phonon coupling models \cite{eph1,eph2,eph3}. 
Using Eqs. (\ref{S1}) and (\ref{S2}), our starting Hamiltonian, $H_{\rm eph}$, is written as 
\begin{align}
H_{\rm eph} &\equiv H_{t} + H_{\rm Coulomb} + H_{\rm EIMV}, \label{S6} \\
H_{\rm EIMV} &= -\sum_{\alpha,j}S_{\alpha}x_{j}^{\alpha}n_{j} + \sum_{\alpha,j}\frac{S_{\alpha}}{2}\left(x_{j}^{\alpha}\right)^{2}, \label{S7} 
\end{align}
where $x_{j}^{\alpha}$ 
denotes the dimensionless reference frame of molecular vibration mode $\alpha$ at the $j$-th site. For a certain vibration 
mode, $\alpha$, we represent $g_{\alpha}$ as an EIMV coupling constant and $\Omega_{\alpha}$ as a frequency of a molecular 
vibration, respectively. Then $S_{\alpha} = 2g_{\alpha}^{2}/\Omega_{\alpha}$. 
Here, we introduce the mean fields of charge disproportion $\delta_{\rm co}$ and amplitude of the molecular vibration 
$x_{0}^{\alpha}$ as follows. 
\begin{align}
\langle\psi_{0}|n_{j}|\psi_{0}\rangle &= \frac{1}{2}+(-1)^{j-1}\frac{\delta_{\rm co}}{2}, \label{S8} \\
x_{j}^{\alpha} &= (-1)^{j-1}x_{0}^{\alpha}. \label{S9} 
\end{align}
After considering the appropriate constant energy shift and comparing $H_{\rm eph}$ in 
Eq. (\ref{S6}) 
with $H_{t}+H_{\rm Coulomb}+H_{\rm eff}^{\rm even}$ (see Eqs. (\ref{S1}), (\ref{S2}), and (\ref{S3})), which is the same as $H$ in 
Eq. (\ref{eqs1}) with $V_{\rm edge}=0$, 
\begin{equation}
V_{\rm eff} = \sum_{\alpha}S_{\alpha}\delta_{\rm co}
\label{S10} 
\end{equation}
is derived by the Hellmann--Feynman theorem. In the charge-ordered (CO) ground state, because a finite $\delta_{\rm co}$ 
deforms the molecular orbitals associated with $n_{j}$ for each site, 
$V_{\rm eff}$ in Eq. (\ref{S10})
can be interpreted as an effective potential representing such deformation. 
\par
To simplify the problem, we select a single molecular vibration mode, 
$\beta$ ($S_{\alpha}=0, x_{j}^{\alpha}=0$ for $\alpha\neq\beta$), and define $S_{\beta}\equiv S$, $x_{j}^{\beta}\equiv x_{0}$. 
Then, we can derive $x_{0}=\delta_{\rm co}/2$ similarly to as discussed above. 
$\delta_{\rm co}$ for a fixed $S/t_{2}$ is determined by minimizing ground-state energy $E_{0}^{\rm EIMV}$ written as 
\begin{equation}
E_{0}^{\rm EIMV} = \langle\psi_{0}|H_{\rm eph}|\psi_{0}\rangle \quad (x_{0}=\delta_{\rm co}/2). 
\label{S11}
\end{equation}
Within the framework of the ED calculation under the PBC, the results of $\delta_{\rm co}$ for $N_{s}=12$, 16, 20 are shown in 
Fig. \ref{sfig3}. To avoid the finite size effect, we use the result for $N_{s}=20$, which is the largest system 
size in our calculations, and estimate $V_{\rm eff}$. 
\par
In a recent experiment \cite{Sawa}, because $\delta_{\rm co}=0.40$ was observed in the CO ground state 
at 30 K, $V_{\rm eff} = S\delta_{\rm co} \sim 0.116t_{2}\equiv V_{\rm eff}^{\rm ED}$ could be estimated at 
$\delta_{\rm co}=0.40$ by using value $S/t_{2}=0.291$ for $N_{s}=20$ at point A, as shown in 
Fig. \ref{sfig3}. 
The estimated value of $V_{\rm eff}^{\rm ED}$ is close to $V_{\rm eff}=0.086t_{2}$. 
\par
Apart from the structural similarity to the first-order phase transition seen in 
Fig. \ref{sfig3}, 
value $S/t_{2}$ vanishes with $\delta_{\rm co}$ of $N_{s}=16$ 
and $N_{s}=20$ is quantitatively regarded as practically unchanged. Therefore, our ED calculations presented in the next subsection, 
focus on $N_{s}=16$. 
\par
Here, we briefly comment on the origin of the first-order phase transition noted in 
Fig. \ref{sfig3}. 
According to the observed $T$-$P$ phase diagram of (TMTTF)$_{2}$PF$_{6}$ \cite{Gex2}, the spin-Peierls (SP) phase 
should be the ground state at $T=0$ and, in general, it should have both 
$4k_{\rm F}$-charge density wave (CDW) and 
$2k_{\rm F}$-spin density wave (SDW) instabilities \cite{INST}, where $k_{\rm F}$ denotes a Fermi wave number. 
In this subsection, we only set CO mean field $\delta_{\rm co}$ that has a $4k_{\rm F}$ instability and do not treat the 
$2k_{\rm F}$ instability of an antiferromagnetic order appearing in the SP state. 
Consequently, a tetrameric model should be considered for ensuring the second-order phase transition of 
$\delta_{\rm co}$, which is our future work. 
%
%
\subsection{Dimer--Mott state in the excited state}
\label{AppdxC}
\par
\begin{figure}[t]
\begin{center}
\includegraphics[width=10cm,keepaspectratio]{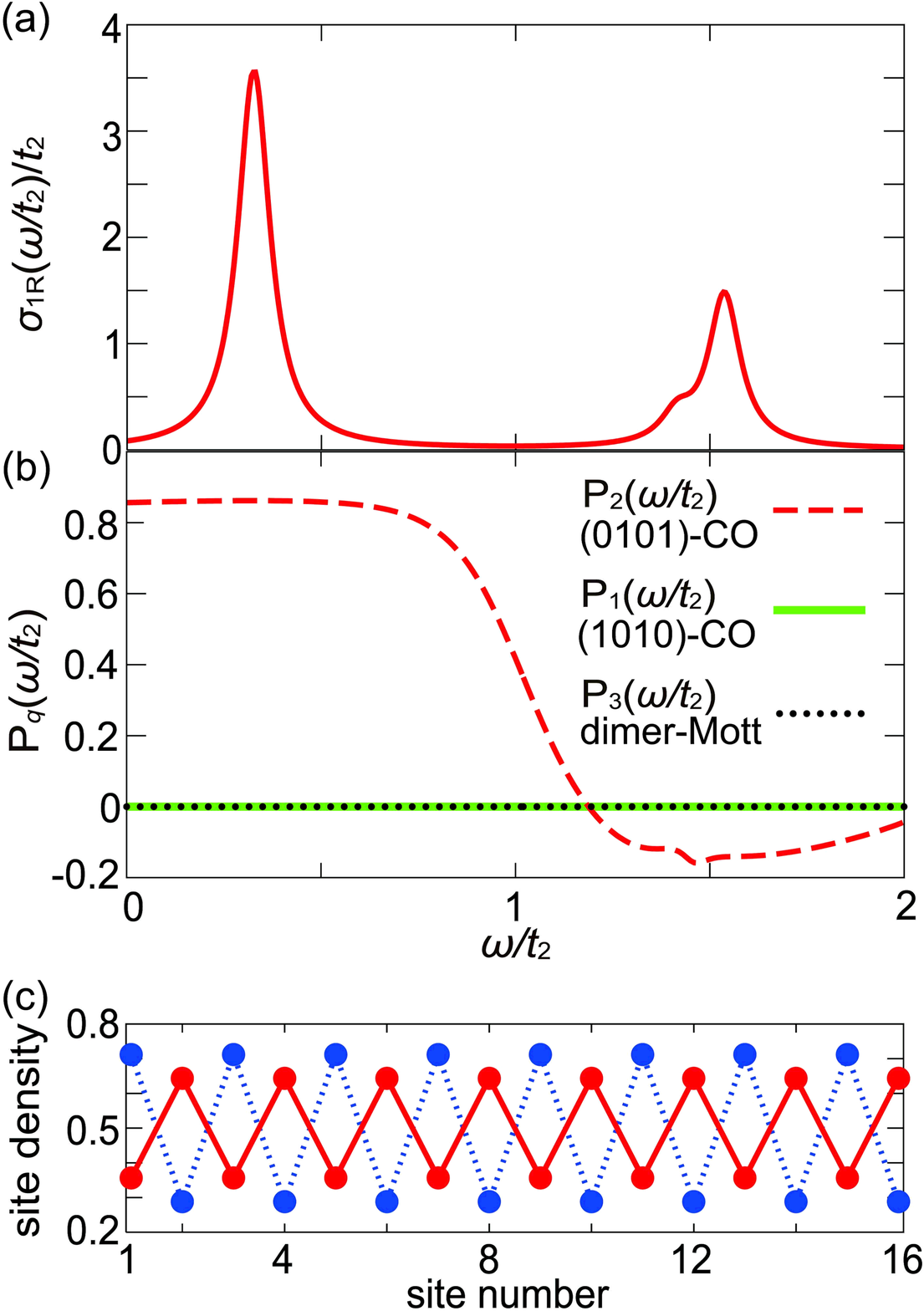}
\caption{Calculations by the ED method under the PBC for $N_{s}=16, \eta/t_{2}=0.05$. 
(a) Computed $\sigma_{\rm 1R}(\omega/t_{2})$ of $H_{1}$ in Eq. (\ref{S12}). 
(b) Projections of a state, $|\Phi_{1}(\omega)\rangle$, onto the ground states at different physical phases 
$|\Psi_{1}\rangle$ ((1010)-CO), $|\Psi_{2}\rangle$ ((0101)-CO), and $|\Psi_{3}\rangle$ (dimer--Mott). 
(c) Site densities. The solid line connected with solid circles describes the site density of the photoexcited 
state at the first peak of $\sigma_{\rm 1R}(\omega/t_{2})$ shown in (a). For comparison, the site density of the 
ground state is displayed as a dotted line with filled circles. 
}
\label{sfig4}
\end{center}
\end{figure}
\par
\par
In this subsection, we investigate the relationship between the dimer--Mott (DM) state and the photoexcited state by means of 
the ED method under the PBC. All the calculations are conducted at $N_{s}=16$ as discussed in the 
previous subsection. In addition, we also inquire regarding the existence of the polarization (P)-inverted CO state 
in the photoexcited state for comparison. For this purpose, using Eqs. (\ref{S1}), (\ref{S2}), and (\ref{S3}), we 
introduce three Hamiltonians defined as 
\begin{align}
H_{1}&\equiv H_{t}+H_{\rm Coulomb}+H_{\rm eff}^{\rm even}, \label{S12} \\
H_{2}&\equiv H_{t}+H_{\rm Coulomb}+H_{\rm eff}^{\rm odd}, \label{S13} \\
H_{3}&\equiv H_{t}+H_{\rm Coulomb}(V=0). \label{S14}
\end{align}
From the discussions in previous sections, because $H_{1}$ in Eq. (\ref{S12}) corresponds to original Hamiltonian 
$H$ with $V_{\rm edge}=0$ in Eq. (\ref{eqs1}), the ground state of $H_{1}$ in Eq. (\ref{S12}) defined as $|\Psi_{1}\rangle$ 
is rich in charges on every odd site and we symbolically represent this as 
``(1010)-CO.'' 
In contrast, defining $|\Psi_{2}\rangle$ as the ground state of $H_{2}$ in Eq. (\ref{S13}) and $|\Psi_{2}\rangle$ as having rich 
charges on each even site, which we symbolically represent as 
``(0101)-CO.'' 
This state can be ascribed as a P-inverted CO state with respect to $|\Psi_{1}\rangle$. 
The DM ground state, $|\Psi_{3}\rangle$, considers the ground state of $H_{3}$ in Eq. (\ref{S14}), 
and we simply refer $|\Psi_{3}\rangle$ as a 
``dimer--Mott.'' 
Although some theoretical works have revealed the phase diagram of the ground state in 1D quarter-filled 
Hamiltonian $H_{t}+H_{\rm Coulomb}$ (see Eqs. (\ref{S1}) and (\ref{S2})) and the parameter region of 
the DM phase with non-vanishing $U$ and $V$ \cite{Ogata,Suzumura}, we particularly choose $V=0$ in $H_{3}$ (Eq. (\ref{S14})) 
for completely neglecting the charge-ordering effects originating from $V\neq 0$. 
\par
We calculate a photoexcited state with given photon energy $\omega$ and $J$ in Eq. (\ref{eqs5}) as 
\begin{equation}
|\Phi_{1}(\omega)\rangle\equiv \frac{1}{\mathcal{N}_{1}}\frac{\eta}{(\omega-E_{1}+H_{1})^{2}+\eta^{2}}J|\Psi_{1}\rangle, 
\label{S15}
\end{equation}
where $E_{1}$ represents the ground-state energy of $H_{1}$ in Eq. (\ref{S12}). $\mathcal{N}_{1}$ is determined by satisfying 
$1=\langle\Phi_{1}(\omega)|\Phi_{1}(\omega)\rangle$. In addition to this, we also calculate 
\begin{equation}
{\rm P}_{q}(\omega)\equiv\langle\Psi_{q}|\Phi_{1}(\omega)\rangle \quad (q=1, 2, 3) 
\label{S16}
\end{equation}
which denotes the characteristic quantities for qualitatively estimating the mixing degrees 
of the different ground states $|\Psi_{q=1,2,3}\rangle$ with respect to $|\Phi_{1}(\omega)\rangle$. 
\par
$\sigma_{\rm 1R}(\omega/t_{2})$ with $\mathcal{H}=H_{1}, E_{0}=E_{1}$, and $|\psi_{0}\rangle=|\Psi_{1}\rangle$ in Eq. (\ref{S5}) 
is shown in 
Fig. \ref{sfig4}(a). 
Our calculations of ${\rm P}_{q=1,2,3}(\omega)$ are displayed 
in Fig. \ref{sfig4}(b). 
$V_{\rm eff}/t_{2}=0.086$ is chosen for the calculations presented here. This values also allow discussing 
the same photoexcitation calculated under the OBC in the previous sections except for finite size effects. 
As it can be seen, in addition to the obvious result of ${\rm P}_{1}(\omega)=0$, 
it is clear that ${\rm P}_{2}(\omega)\neq 0$ and ${\rm P}_{3}(\omega)=0$. In particular, ${\rm P}_{2}(\omega)\sim 1$ can be seen 
around the first peak of $\sigma_{\rm 1R}(\omega/t_{2})$. Therefore, the photoexcited state at the first peak is highly inclusive 
of the P-inverted state, $|\Psi_{2}\rangle$ ((0101)-CO), regarding the $|\Psi_{1}\rangle$ ((1010)-CO) ground state, but it is exclusive 
of the DM state, $|\Psi_{3}\rangle$. Compared to the site density of the ground state $\langle\Psi_{1}|n_{j}|\Psi_{1}\rangle$, 
the enhancement of the ``(0101)-CO'' photoexcited state at the first peak is consistent with site density 
at the peak $\langle\Phi_{1}(\omega)|n_{j}|\Phi_{1}(\omega)\rangle$ shown in 
Fig. \ref{sfig4}(c). 
%
%

\end{document}